# On the Optimal Interaction Range for Multi-Agent Systems Under Adversarial Attack

Saad J Saleh

*Abstract*— Consider a consensus-driven multi-agent dynamic system. The interaction range, which defines the set of neighbors for each agent, plays a key role in influencing connectivity of the underlying network. In this paper, we assume the system is under attack by a predator and explore the question of finding the optimal interaction range that facilitates the most-efficient escape trajectories for the group of agents. We find that for many cases of interest the optimal interaction range is one that forces the network to break up into a handful of disconnected graphs, each containing a subset of agents, thus outperforming the two extreme cases corresponding to fully-connected and fully-disconnected networks. In other words, the results indicate that some connectivity among the agents is helpful because information is effectively transmitted from the agents closest to the predator to others slightly farther away, but also that too much connectivity can be detrimental to the agility of the group, thus hampering efficient and rapid escape

## I. INTRODUCTION

The desire to understand and control the dynamics of networked self-propelled particles has motivated significant research efforts in a variety of disciplines over the last three decades. Examples include the fields of biological physics [1]-[2], distributed computing [3], mobile robots [4], networked sensors [5], formation control [6]-[7], space rendezvous problems [8], and opinion dynamics [9]. A large body of results concerning various dynamical aspects, such as stability, convergence speed, robustness, and leader tracking accuracy can be found in several literature surveys [10]-[14].

One common theme in the results mentioned above concerns the notion of network connectivity. Although there are various subtle ways of defining system connectivity (as a function of time), some more restrictive than others, it is perhaps fair to say that the majority of early consensus and cooperation results associated with multi-agent systems assume the underlying network remains "largely connected" as it evolves in time. An interesting exception to this rule of thumb can be found in the biological physics literature. For example, in the Vicsek model for simulating starling flocks introduced in [1], the authors highlight parameter-dependent phase transitions that are often linked to various connectivity aspects of the overall system. Another example can be found in the literature on opinion dynamics, where disagreement [15] or antagonism [16] among agents is often of interest.

Naturally, preserving connectivity during the system's time evolution is a sensical constraint to impose for most applications of networked multi-agent systems such as consensus, cooperation, and leader-following. However, for certain applications and considerations, such as effective escape mechanisms when the agents are under attack, the ability to break up rapidly into disconnected groups may offer certain advantages – an example being the case of a flock of birds escaping an approaching predator. This consideration provides one of the main motivations for this paper. Note that the problem of investigating evasive strategies for multi-agent systems in the presence of a pursuer has been explored by several researchers in recent years [17]-[20]. In contrast to these studies, however, we focus here on network connectivity as a potentially critical aspect for successful evasion or escape.

There are several system parameters that influence connectivity and its subsequent impact on potentially advantageous behavior for systems under attack. In this paper, we focus on one particular parameter – namely, the interaction range, which defines the radius of influence around each agent. It can readily be shown, and is intuitively obvious, that the interaction range strongly impacts connectivity of the network: the larger the range, the more likely connectivity is preserved. For relatively small interaction ranges, the underlying network "breaks up" into a number of disconnected graphs. Now, assume the agents to be under attack by a predator, so that their dynamic behavior is governed by a superposition of two forces, one being a typical consensus algorithm and the other being a repulsive force that facilitates escape from the predator. Under this scenario, one can define an objective function that represents escape efficiency and search for the optimal interaction range that maximizes this function. We address this problem first for the simple case where each agent is represented by a scalar (e.g., representing direction of movement) and later for the more realistic case where each agent is represented by a 6-dimensional vector containing the coordinates of its position and velocity. We refer to the first instance as the 1D case and the second as the 3D case as these are the dimensions of the associated configuration spaces. We find that in many cases involving both the 1D and 3D scenarios the optimal interaction range lies in some "middle ground" between the two extremes corresponding to a large range (fully connected networks) and a small range (fully disconnected networks). This is an interesting result as it reinforces the intuition that some connectivity among the agents is helpful because information is effectively transmitted from the agents closest to the predator to others slightly farther away, but also that too much connectivity can be detrimental as the critical repulsive force gets masked by the stronger consensus force, thus hampering efficient and rapid escape.

S. J. Saleh is with the Department of Electrical and Computer Engineering, Rice University, Houston, TX, USA. (e-mail: Saad.Saleh@rice.edu).

The escape efficiency problem can be addressed within the context of the system's transient response or, alternatively, its steady-state response. It is also possible to formulate objective functions that combine both responses. In this paper, after providing a mathematical description of the system in Section II, we present a few theoretical results relevant to the steady-state behavior of the 1D case in Section III, which in turn allow us to introduce the desired optimization problem. We address this problem numerically in Section IV. In Section V, we shift focus to the 3D case, reformulate the problem in terms of the system's transient response, and present numerical solutions for this more general scenario. Finally, we provide our conclusions in Section VI.

## II. PROBLEM FORMULATION

Consider a discrete-time system consisting of $n$ agents labelled $1, \ldots, n$, with the time-dependent state of the $i$th agent denoted by $x_i(t) \in \mathbb{R}$. Let $\mathcal{N}_i(t)$ denote the set of neighbors for agent $i$ at time $t$, defined by

$$j \in \mathcal{N}_i(t) \Leftrightarrow |x_i(t) - x_j(t)| \leq \rho \quad i,j = 1, \ldots, n$$

with the real constant $\rho \geq 0$ being a parameter that characterizes the interaction range for the system. In this paper, we restrict $t$ to the set of non-negative integers $\{0,1,2,\ldots\}$ and focus on the classic consensus problem in which the state of each agent evolves in time according to the rule

$$x_i(t+1) = \frac{1}{|\mathcal{N}_i(t)|} \sum_{j \in \mathcal{N}_i(t)} x_j(t) \quad (1)$$

$$x_i(0) = x_{i,0}$$

where $|\mathcal{N}_i(t)|$ denotes the cardinality of $\mathcal{N}_i(t)$, and $x_{i,0}$ is the initial state of the $i$th agent. As is customary for this type of setup, we view the system at each time $t$ as an (undirected) graph $\mathcal{G}(t) = (\mathcal{V}, \mathcal{E}(t))$ where the set of graph nodes $\mathcal{V} = \{1, \ldots, n\}$ consists of the agents' labels, and the set of graph edges $\mathcal{E}(t)$ is given by the neighbors rule

$$(i,j) \in \mathcal{E}(t) \Leftrightarrow j \in \mathcal{N}_i(t) \Leftrightarrow i \in \mathcal{N}_j(t)$$

With this setup, the dynamic system given by (1) can be compactly written (e.g., see [10]) as

$$x(t+1) = P(t)x(t), \quad (2)$$

$$x(0) = x_0$$

with $x(t) \doteq [x_1(t) \ldots x_n(t)]^T$, $x_0 \doteq [x_{1,0} \cdots x_{n,0}]^T$ and $P(t)$, the graph's Perron matrix, given by

$$P(t) = I - \Delta^{-1}(t)L(t),$$

where $I$ is the $n \times n$ identity matrix, $\Delta(t)$ is the graph's Degree matrix, and $L(t)$ is the graph's Laplacian matrix.

One significant advantage of using the graph-theoretic formulation of (2) is afforded by the wealth of information accumulated over the last few decades regarding the spectral properties of graph Laplacian matrices which can be readily employed to investigate convergence properties of this dynamic system. Furthermore, this setup also facilitates viewing the evolving network as a simple time-varying feedback system as shown in Figure 1.

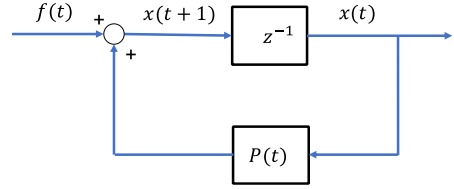

Figure 1: System Block Diagram

The formulation given above in (1) and (2) is clearly limited to autonomous systems with no external input; i.e., with $f(t) = 0$ in Figure 1. There is a wealth of recent literature aimed at extending this setup to forced systems where the external input $f(t)$ may, for example, represent a signal to be tracked by the agents; e.g., leader-following examples. In this paper, we are also interested in incorporating external input signals but our main objective is to consider predators to be avoided, as opposed to leaders to be followed. To this end, we consider several examples of predator-related $f(t)$ for both the 1D case (in Section III) and 3D case (in Section V).

## III. THEORETICAL RESULTS: 1D CASE

We start by considering the unforced system given in (2). It is well-known (e.g., see [21]) that if the graph $\mathcal{G}(t)$ associated with $x(t)$ remains connected for all $t \geq 0$, then $x(t)$ converges to a steady-state vector given by $\alpha \mathbf{1}$, where $\alpha \in \mathbb{R}$ is the average value of the components of $x(0)$ and $\mathbf{1} = [1 \cdots 1]^T$. Our first goal is to extend this result to the case of unconnected graphs. To this end, we first need the following lemma, which is a variant on a similar result by Dittmer [22].

*Lemma 1*: Consider the $n$-dimensional system given by (2), with some arbitrary $x(0) \in \mathbb{R}^n$. Let $\mathcal{G}(t)$ be the graph associated with $x(t)$, and let $m(t) \in \mathbb{Z}^+$ be the number of connected components of $\mathcal{G}(t)$. Then, under the consensus algorithm given by (1),

1) $m(t+1) \geq m(t)$ for all $t \geq 0$,
2) there exists a positive integer $m^*$, with $1 \leq m^* \leq n$, such that

$$\lim_{t \to \infty} m(t) = m^*.$$

*Proof of Lemma 1*: It is clear that if $\mathcal{G}(t)$ remains connected for all $t \geq 0$, then the lemma is satisfied with

$$m(t+1) = m(t) = m^* = 1 \quad \text{for all } t \geq 0.$$

Hence, we need only consider the case where connectedness of $\mathcal{G}(t)$ is broken at some $t \geq 0$. Let $t_b$ be the smallest $t$ at which this occurs; i.e.,

$$t_b = \min_{t \geq 0}\{t : m(t) > 1\}.$$

Thus $m(t_b) > 1$ is the number of connected components of $\mathcal{G}(t_b)$. Now, let $\mathcal{G}_1(t_b), \ldots, \mathcal{G}_{m(t_b)}(t_b)$ be the connected components of $\mathcal{G}(t_b)$, and choose $\mathcal{G}_k(t_b)$ and $\mathcal{G}_l(t_b)$ to be any two members of this collection. Furthermore, let $x^k(t_b) \in \mathbb{R}^{|\mathcal{G}_k(t_b)|}$ be the sub-vector of $x(t_b)$ containing the values taken by the agents in $\mathcal{G}_k(t_b)$ and $x^l(t_b) \in \mathbb{R}^{|\mathcal{G}_l(t_b)|}$ be the sub-vector of $x(t_b)$ containing the values taken by the agents in $\mathcal{G}_l(t_b)$. Now, define the distance between $\mathcal{G}_k(t_b)$ and $\mathcal{G}_l(t_b)$ as

$$d(\mathcal{G}_k(t_b), \mathcal{G}_l(t_b)) \doteq \min_{\substack{i=1,\ldots,|\mathcal{G}_k(t_b)| \\ j=1,\ldots,|\mathcal{G}_l(t_b)|}} \left| x_i^k(t_b) - x_j^l(t_b) \right|$$

We claim that

$$\rho \leq d(\mathcal{G}_k(t_b), \mathcal{G}_l(t_b)) \leq d(\mathcal{G}_k(t_b+1), \mathcal{G}_l(t_b+1)) \quad (3)$$

The first inequality follows directly as a result of $\mathcal{G}_k(t_b)$ and $\mathcal{G}_l(t_b)$ being disconnected. To demonstrate the second inequality, let

$$C^k(t_b) \doteq \left[ x_{min}^k(t_b), x_{max}^k(t_b) \right] \subset \mathbb{R}$$

and

$$C^l(t_b) \doteq \left[ x_{min}^l(t_b), x_{max}^l(t_b) \right] \subset \mathbb{R}$$

be the convex hulls of the elements of $x^k(t_b)$ and $x^l(t_b)$, respectively. Note that $C^k(t_b) \cap C^l(t_b) = \emptyset$.

Now, since for any agent $i \in \mathcal{G}_k(t_b)$, $x_i^k(t_b+1)$ is a convex combination of the elements of $x^k(t_b)$ (see Equation (1)) and likewise for any agent $j \in \mathcal{G}_l(t_b)$, $x_j^l(t_b+1)$ is a convex combination of the elements of $x^l(t_b)$, it follows that

$$x_i^k(t_b+1) \in C^k(t_b+1) \subseteq C^k(t_b)$$

and

$$x_j^l(t_b+1) \in C^l(t_b+1) \subseteq C^l(t_b)$$

from which the second inequality in (3) follows. This shows that if $\mathcal{G}_k(t_b)$ and $\mathcal{G}_l(t_b)$ are disconnected, then so are $\mathcal{G}_k(t_b+1)$ and $\mathcal{G}_l(t_b+1)$, which in turn demonstrates that $m(t)$ is a non-decreasing function of $t$, proving the first claim of the lemma.

Clearly, the number of connected components cannot exceed the number of agents $n$. Hence, $m(t)$ is a non-decreasing function of time that is bounded above by $n$, which proves the second claim. ∎

Before proceeding to the first main result, we need the following definitions.

*Definition 1*: Let $m^*$ be the number of connected components of system (2) at steady state, as in *Lemma 1*. Let $\mathcal{G}_i^*, i = 1, \ldots, m^*$ be the associated components (sub-graphs of $\mathcal{G}$) at steady state, with $|\mathcal{G}_i^*|$ denoting the number of agents included in $\mathcal{G}_i^*$. Let $t_c$ be the smallest $t$ at which $m(t) = m^*$; i.e.,

$$t_c = \min_{t \geq 0} \{t : m(t) = m^*\}. \quad (4)$$

Consider the vectors $x^1(t_c), \ldots, x^{m^*}(t_c)$ such that each $x^i(t_c) \in \mathbb{R}^{|\mathcal{G}_i^*|}$ ($i = 1, \ldots, m^*$) contains the states of all agents in $\mathcal{G}_i^*$ at time $t_c$. Now, define the scalars

$$\alpha_i = \frac{1}{|\mathcal{G}_i^*|} \sum_{j=1}^{|\mathcal{G}_i^*|} x_j^i(t_c),$$

and the vectors

$$\mathbf{1}^i = [1 \quad \cdots \quad 1]^T \in \mathbb{R}^{|\mathcal{G}_i^*|}.$$

Furthermore, order the $\alpha_i$s so that $\alpha_1 \leq \cdots \leq \alpha_{m^*}$ and define the steady state vector

$$x^* = \begin{bmatrix} \alpha_1 \mathbf{1}^1 \\ \vdots \\ \alpha_{m^*} \mathbf{1}^{m^*} \end{bmatrix} \doteq \begin{bmatrix} x^1 \\ \vdots \\ x^{m^*} \end{bmatrix}. \quad (5)$$

Finally, define the steady state set $\mathcal{T}$ as

$$\mathcal{T} = \{Qx^* : Q \in \mathbb{R}^{n \times n} \text{ is a permutation matrix}\}. \quad (6)$$

We are now ready to state our first theorem.

*Theorem 1*: The dynamic system (2) converges to a steady state vector $x_{ss} \in \mathcal{T}$, where $\mathcal{T}$ is defined in (6).

*Proof of Theorem 1*: Consider $t_c$ as defined in (4). Lemma 1 shows that for all $t \geq t_c$, the system consists of $m^*$ uncoupled sub-systems. Moreover, each of the sub-systems $1, \ldots, m^*$ remains connected for all $t \geq t_c$. Hence, for any $k \in \{1, \ldots, m^*\}$ the sub-system labelled $k$ converges (e.g., see [21]) to the corresponding steady-state sub-vector $x^k$ defined in (5). Now, if the elements of $x(t_c)$ were ordered such that $x_i(t_c) \leq x_{i+1}(t_c)$, then the argument above shows that $x(t)$ converges to $x^*$ as defined in (5). However, since node labeling is arbitrary, $x(t)$ will in fact converge to a permutation of $x^*$, depending on the arbitrary labeling scheme; i.e., it converges to a vector $x_{ss} \in \mathcal{T}$, as defined in (6). ∎

We are now in a position to expand the analysis to the case where the multi-agent network is under attack by predators. For simplicity, we restrict the discussion to one predator $p$ which is assumed to have a stationary position at $x_p \in \mathbb{R}$ for all $t \geq 0$. Furthermore, we assume that presence of the predator can impact behavior of only those agents that are within its own range of influence $\rho_p \geq 0$, which may be different from the agents' interaction range $\rho$. This predator-induced behavior of the agents is typically a repulsive "escape" force. We also introduce a real constant $s > 0$ to characterize the strength of repulsion due to the predator's presence. To make these ideas concrete, we modify Equation (1) as:

$$x_i(t+1) = \frac{1}{|\mathcal{N}_i(t)|} \sum_{j \in \mathcal{N}_i(t)} x_j(t) + f(x_i(t)), \quad (7)$$

where $f(x_i(t))$ is the repulsive force experienced by agent $i$ due to being within the predator's range of influence. This function can take on several forms. One simple example is

$$f(x_i(t)) = \begin{cases} s \, \text{sign}(x_i(t) - x_p) & \text{if } |x_i(t) - x_p| < \rho_p \\ 0 & \text{if } |x_i(t) - x_p| \geq \rho_p \end{cases} \quad (8)$$

where

$$\text{sign}(w) = \begin{cases} +1 & \text{if } w > 0 \\ 0 & \text{if } w = 0 \\ -1 & \text{if } w < 0 \end{cases}$$

Another example is

$$f(x_i(t)) = \begin{cases} x_p - x_i(t) + s \, \rho_p \, \text{sign}(x_i(t) - x_p) & \text{if } |x_i(t) - x_p| < \rho_p \\ 0 & \text{if } |x_i(t) - x_p| \geq \rho_p \end{cases}$$
(9)

Both of these examples are shown in Figure 2.

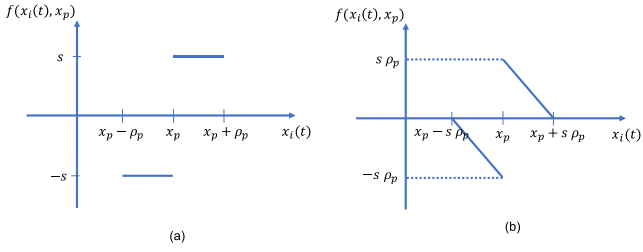

Figure 2: Two Examples of Predator Repulsive Force

With this setup, we are now ready to state our second theorem.

*Theorem 2*: Consider the dynamic system (7)-(8) under threat by one predator that remains fixed at one position $x_p \in \mathbb{R}$ for all $t \geq 0$. Assume $\rho_p \geq \rho$. Then, there exists a repulsive strength $s^* > 0$ such that the system converges to a steady state value $x_{ss}$ for all $s \geq s^*$

$$\lim_{t \to \infty} x(t) = x_{ss}$$

where $x_{ss}$ is a function of $x(0)$, $x_p$, $s$, $\rho$, and $\rho_p$

*Proof of Theorem 2*: First we show there exists an $s^* > 0$ and time $t^* \geq 0$ such that for all states $x_i(t^*)$, the following inequality holds

$$|x_i(t^* + 1) - x_p| > \rho_p. \quad (10)$$

To see this, pick any $t^* \geq 0$ and any arbitrary agent whose state is given by $x_i(t^*)$, and consider 2 cases:

Case 1: $x_i(t^*) \geq x_p$

Notice first that since the expression

$$(1/|\mathcal{N}_i(t^*)|) \sum_{j \in \mathcal{N}_i(t^*)} x_j(t^*)$$

is a convex combination of $x_i(t^*)$ and its neighbors, it follows that it's bounded above and below as

$$x_i(t^*) - \rho \leq \frac{1}{|\mathcal{N}_i(t^*)|} \sum_{j \in \mathcal{N}_i(t^*)} x_j(t^*) \leq x_i(t^*) + \rho$$

Therefore,

$$x_i(t^* + 1) - x_p = \frac{1}{|\mathcal{N}_i(t^*)|} \sum_{j \in \mathcal{N}_i(t)} x_j(t^*) + s - x_p$$

$$\geq x_i(t^*) - \rho + s - x_p$$

$$\geq x_p - \rho + s - x_p$$

$$= s - \rho$$

Now, choose $s^* = \rho_p + \rho$. This choice ensures satisfaction of (10) for Case 1.

Case 2: $x_i(t^*) < x_p$

Nearly identical arguments to those given for Case 1 demonstrate that the choice $s^* = \rho_p + \rho$ leads to

$$x_i(t^* + 1) - x_p \leq -\rho_p$$

which ensures satisfaction of (10) for Case 2.

Now, since $x_i(t^*)$ was chosen arbitrarily, it follows that (10) holds for all agents under this choice of $s^*$.

The arguments above show that the choice $s^* = \rho_p + \rho$ separates the agents' positions at $t^* + 1$ into two sets $x^u(t^* + 1)$ and $x^w(t^* + 1)$ such that

$$x^u(t^* + 1) = \{x_i(t^* + 1): x_i(t^* + 1) - x_p > \rho_p\}$$

$$x^w(t^* + 1) = \{x_i(t^* + 1): x_i(t^* + 1) - x_p < -\rho_p\}.$$

We claim that for all $t \geq t^* + 1$, the system remains predator-free and will thus converge to a steady-state value as predicted by Theorem 1. To see this, let

$$C^u \doteq [x^u_{min}(t^* + 1), x^u_{max}(t^* + 1)] \subset \mathbb{R}$$

and

$$C^w \doteq [x^w_{min}(t^* + 1), x^w_{max}(t^* + 1)] \subset \mathbb{R}$$

be the convex hulls of the elements of $x^u(t^* + 1)$ and $x^w(t^* + 1)$, respectively. Now, since

$$d(C^u, C^w) \doteq x^u_{min}(t^* + 1) - x^w_{max}(t^* + 1)$$

$$\geq \rho_p$$

$$\geq \rho$$

and since the expression $\frac{1}{|\mathcal{N}_i(t)|} \sum_{j \in \mathcal{N}_i(t)} x_j(t)$ is a convex combination of $x_i(t)$ and its neighbors, it follows that (see Equation (7))

$$x_i(t^* + 1) \in x^u(t^* + 1) \Rightarrow x_i(t) \in C^u \text{ for all } t \geq t^* + 1$$

and

$$x_i(t^* + 1) \in x^w(t^* + 1) \Rightarrow x_i(t) \in C^w \text{ for all } t \geq t^* + 1.$$

In other words, the agents with positions in $C^u$ at $t = t^* + 1$ will not be influenced by the predator or by any agent whose position is in $C^w$ for all $t \geq t^* + 1$, and likewise the agents with positions in $C^w$ at $t = t^* + 1$ will not be influenced by the predator or by any agent whose position is in $C^u$ for all $t \geq t^* + 1$. ∎

*Remark:* Theorem 2 was proven for the repulsive force given in (8) but it can be shown to hold for a large family of repulsive forces, including for example the one given in (9).

We are now in a position to define an objective function for the optimization problem at hand. Escape efficiency can be addressed either from a transient response point of view or from a steady-state point of view, or some combination of both. In this section, we focus on the steady-state response. Theorem 2 shows that for a sufficiently large repulsive force they system converges to a steady state, which allows us to make the following definition.

*Definition 2*. Consider the system (7)-(8) and assume the conditions of Theorem 2 hold, particularly that the repulsive strength $s$ is sufficiently large to guarantee the existence of a steady state $x_{ss}$. Then, for any interaction range $\rho \geq 0$, let the time-dependent escape distance be

$$d_\rho(t) \doteq \min_{i \in \{1,\ldots,n\}} |x_i(t) - x_p|$$

Now, we define the steady-state escape distance to be

$$d_{ss}(\rho) = \lim_{t \to \infty} d_\rho(t)$$

With the setup above, the main problem can be stated as:

*Problem 1.* For a system under threat by one predator that remains fixed at one position $x_p$ for all $t \geq 0$, find the optimal interaction distance $\rho^*$ for which

$$\rho^* = \max_{\rho \geq 0} d_{ss}(\rho)$$

In the next section we explore numerical solutions for Problem 1.

## IV. NUMERICAL RESULTS: 1D CASE

To shed more light on the results obtained in the previous section, we first simulate the time evolution of system (1) with $n = 100$, and $x(0)$ being a random vector where each $x_i(0)$ is uniformly distributed on [0,1]. Results for one realization corresponding to the predator-free case are shown in Figure 3 for $\rho = 1$, 0.2, 0.1, and 0. The $\rho = 1$ case (Figure 3a) represents a classic consensus example with a relatively large range of influence where $x$ converges to $\alpha \mathbf{1}$, with $\alpha$ being the average value of the components of $x(0)$, which is 0.5 in this case. As expected, the network remains connected throughout. The other extreme scenario is represented by the $\rho = 0$ case (Figure 3d) where agents are fully independent of each other and the network remains completely disconnected throughout as 100 independent nodes. The remaining illustrated cases of $\rho = 0.2$ (Figure 3b) and $\rho = 0.1$ (Figure 3c) demonstrate the middle-ground cases where the system reaches a steady state consisting of 2 clusters, and 4 clusters, respectively. In reference to the steady-state vector $x^*$ given in (5), the $\rho = 0.2$ case leads to $m^* = 2$, $\alpha_1 = 0.35$, $\alpha_2 = 0.69$, whereas $\rho = 0.1$ leads to $m^* = 4$, $\alpha_1 = 0.14$, $\alpha_2 = 0.40$, $\alpha_3 = 0.67$, $\alpha_4 = 0.88$.

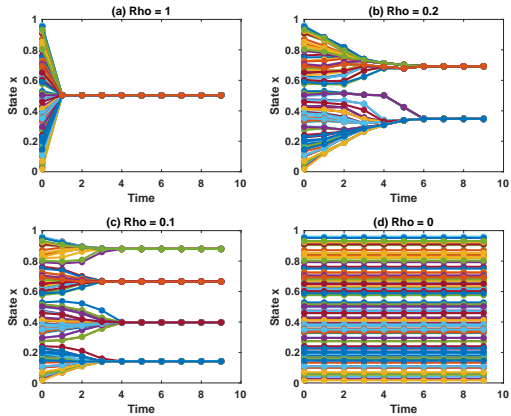

Figure 3: Predator-Free System 1D Case

Next, we re-consider the same system, with the same initial state, but with a predator located at $x_p = 0.6$ for all $t \geq 0$. We fix the predator range of influence at $\rho_p = 0.2$, and adopt Equation (9) as the predator repulsive force, with a strength $s = 2$. With this setup, we repeat the time evolution simulations for the 4 interaction ranges considered in Figure 3 but now with a predator incorporated into the system as outlined above. The results are shown in Figure 4. The steady-state escape distances $d_{ss}(\rho)$ for the 4 cases can be estimated from the figure, where it is apparent that the best response among those 4 cases is attained for $\rho = 0.1$, with $d_{ss}(\rho) = 0.3$.

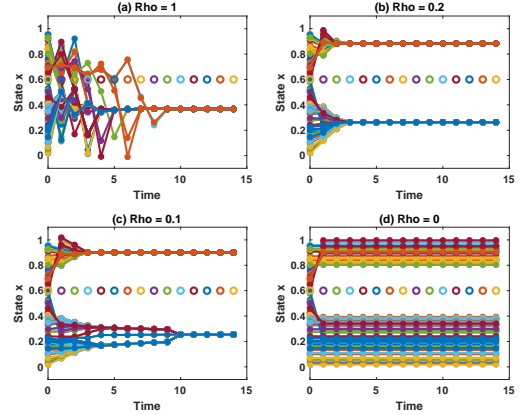

Figure 4: 1D Case with Predator Represented by "o"

The results above are clearly dependent on the particular realization chosen for $x(0)$. To capture the stochastic nature of the system due to randomness of the initial conditions, one can reformulate Problem 1 of the previous section so as to maximize the mean value of steady-state escape distance over a wide range of initial conditions. To this end, we show in Figure 5(a) an estimate of the mean steady-state escape distance as a function of the interaction range $0 \leq \rho \leq 1$, computed as an average of 40 trials using uniformly distributed values of $x_i(0)$ over [0,1]. The predator parameters used for this simulation are the same as before: $x_p = 0.6$ and $s = 2$. The corresponding plot for the number of average clusters at steady state for each interaction range is shown Figure 5(b). It is clear that for this particular set of predator parameters, the optimal interaction range is $\rho = 0.1$, which results in a mean value of 0.3 for $d_{ss}(\rho)$ and corresponds to the system breaking up into 2 or 3 clusters in steady state (the mean value of the number of clusters corresponding to $\rho = 0.1$ in Figure 5(b) is 2.65.)

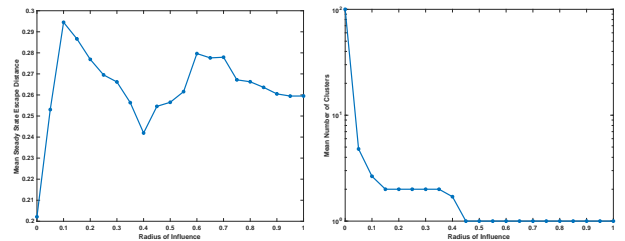

Figure 5: (a) Mean Steady State Escape Distance as a Function of $\rho$. (b) Mean Number of Clusters as a Function of $\rho$

## V. NUMERICAL RESULTS: 3D CASE

To extend our analysis to the case of a 3-dimensional configuration space, we now associate with each agent $i \in \{1, \ldots, n\}$ at every $t \geq 0$ a position vector $r_i(t) \in \mathbb{R}^3$ and a

velocity vector $v_i(t) \in \mathbb{R}^3$ so that the state $x(t) \in \mathbb{R}^{6n}$ is given by

$$x(t) = \begin{bmatrix} r_1(t) \\ \vdots \\ r_n(t) \\ v_1(t) \\ \vdots \\ v_n(t) \end{bmatrix} \doteq \begin{bmatrix} r(t) \\ v(t) \end{bmatrix}$$

As such, the set of neighbors $\mathcal{N}_i(t)$ for the $i$th agent at time $t$ is defined by

$$j \in \mathcal{N}_i(t) \Leftrightarrow \|r_i(t) - r_j(t)\| \leq \rho \qquad i,j = 1, \ldots, n$$

for some appropriate norm $\|.\|$ on $\mathbb{R}^3$, where $\rho \geq 0$ is the interaction range parameter as in the 1D case.

The discretized Newtonian laws of motion for this system are given by

$$\begin{bmatrix} r(t+\Delta t) \\ v(t+\Delta t) \end{bmatrix} = \begin{bmatrix} I & (\Delta t)I \\ 0 & I \end{bmatrix} \begin{bmatrix} r(t) \\ v(t) \end{bmatrix} + \begin{bmatrix} 0 \\ (\Delta t)M^{-1} \end{bmatrix} f(t)$$

where $\Delta t > 0$ is the time increment, $I \in \mathbb{R}^{3n \times 3n}$ and $0 \in \mathbb{R}^{3n \times 3n}$ are the identity and zero matrices, respectively, and $M \in \mathbb{R}^{3n \times 3n}$ is a block diagonal matrix with diagonal entries $M_i \in \mathbb{R}^{3 \times 3}$ given by

$$M_i = m_i \, I_{3\times 3}$$

where $m_i > 0$ is mass of the $i$th agent. For the predator-free case, the forcing function $f(t) \in \mathbb{R}^{3n}$ is chosen so that the consensus algorithm is applied to the direction of the agents' velocities while they maintain constant speed; i.e.,

$$f(t) = \begin{bmatrix} f_1(t) \\ \vdots \\ f_n(t) \end{bmatrix}$$

with $f_i(t) \in \mathbb{R}^3, i \in \{1, \ldots, n\}$ given by

$$f_i(t) = \frac{v_0 \langle v_i(t) \rangle}{\|\langle v_i(t) \rangle\|}$$

(11)

where

$$\langle v_i(t) \rangle = \frac{\sum_{j \in \mathcal{N}_i(t)} v_j(t)}{|\mathcal{N}_i(t)|}$$

and $|\mathcal{N}_i(t)|$ denotes the cardinality of $\mathcal{N}_i(t)$ as before.

To introduce a predator into this system, let $x_p(t) \in \mathbb{R}^6$ represent the predator's state, with

$$x_p(t) = \begin{bmatrix} r_p(t) \\ v_p(t) \end{bmatrix}$$

where $r_p(t) \in \mathbb{R}^3$ and $v_p(t) \in \mathbb{R}^3$ are the predator's position and velocity vectors. We assume the predator's velocity is constant so that its dynamics are described by

$$r_p(t+\Delta t) = r_p(t) + (\Delta t)v_p(t),$$
$$v_p(t+\Delta t) = v_p(t).$$

As in the 1D case, we assume the predator has its own interaction range $\rho_p \geq 0$, which defines its set of neighbors $\mathcal{N}_p(t)$ by

$$j \in \mathcal{N}_p(t) \Leftrightarrow \|r_j(t) - r_p(t)\| \leq \rho_p \qquad i,j = 1, \ldots, n.$$

With this setup, the laws of motion for the agents are modified so that the forcing function $f_i(t)$ is now given by

$$f_i(t) = f_{i,1}(t) + f_{i,2}(t)$$

where the consensus function $f_{i,1}(t)$ is the same as in the predator-free case (Equations (11)) while the escape function $f_{i,2}(t)$ is given by

$$f_{i,2}(t) = \begin{cases} \dfrac{s(r_i(t) - r_p(t))}{\|r_i(t) - r_p(t)\|} & \text{if } i \in \mathcal{N}_p(t) \\ 0 & \text{if } i \notin \mathcal{N}_p(t) \end{cases}$$

where, in analogy to the 1D case, $s > 0$ represents repulsion strength.

In contrast to the 1D case where we focused on the system's steady-state behavior, we consider here the transient response. Specifically, we define the predator distance function $d_{i,\rho}(t)$ for each agent $i \in \{1, \ldots, n\}$ and each interaction range $\rho \geq 0$ as

$$d_{i,\rho}(t) = \|r_i(t) - r_p(t)\|,$$

the average predator distance function as

$$\bar{d}_\rho(t) = \frac{1}{n}\sum_{i=1}^n d_{i,\rho}(t),$$

and search for the optimal $\rho^* \geq 0$ that satisfies

$$\rho^* = \max_{\rho \geq 0} \min_{t \geq 0} \bar{d}_\rho(t). \qquad (12)$$

Alternatively, we may consider the minimum predator distance function

$$\check{d}_\rho(t) = \min_{i \in \{1, \ldots, n\}} d_{i,\rho}(t)$$

and search for the optimal $\rho^* \geq 0$ that satisfies

$$\rho^* = \max_{\rho \geq 0} \min_{t \geq 0} \check{d}_\rho(t). \qquad (13)$$

To shed light on the setup above, we provide next the results of a numerical simulation. To this end, we take $\Delta t = 0.05$ s, assume the number of agents (e.g., birds) to be $n = 300$, and let the mass of each agent be 0.1 kg. For ease of display, we restrict the simulation to a 2-dimensional configuration space and assume the agents are confined initially to a square of size $100^2$ m$^2$; i.e., we let the initial positions $r_i(0)$ for $i \in \{1, \ldots, n\}$ be randomly uniformly distributed over $[0,100] \times [0,100] \times \{0\}$. With regard to velocities, we assume the speed of each agent to be constant at $v_0 = 10$ m/s, and the initial directions are uniformly distributed over $[0,2\pi] \times [0,2\pi] \times \{0\}$. Furthermore, we assume the predator's initial position is $r_p(0) = [-30 \; -30 \; 0]^T$, its initial velocity is $v_p(0) = [10 \; 10 \; 0]^T$, its range of influence is $\rho_p = 30$ m, and its repulsion strength is $s = 10$. We use the Euclidean norm throughout these simulations for measuring distances.

With the setup above, we show in Figure 6 snapshots of the predator and agents positions at $t = 0, 4, 8,$ and 12 s for one realization corresponding to $\rho = 100$ m. We repeat the simulation using the same initial conditions for the cases corresponding to $\rho = 10$ m and $\rho = 0$ m and show the resulting snapshots in Figures 7 and 8, respectively. Qualitatively, these figures hint at the advantage of having short non-zero interaction ranges. In other words, for the $\rho = 10$ m case, break-up of the group network into 12 disconnected graphs seems to facilitate an efficient escape, at least visually, when compared to the $\rho = 100$ m case in which the group remains connected for all $t \geq 0$ and also when compared to the $\rho = 0$ m case in which there is no connectivity at all among the agents. To make these ideas more quantitative, we show in Figure 9(a) the average predator distance function $\bar{d}_\rho(t)$ for $\rho = 100, 10,$ and $0$ m, which demonstrates superiority of the $\rho = 10$ m case among the 3 interaction ranges considered in the sense that it results in a larger average predator distance than the one attained with $\rho = 100$ m and $\rho = 0$ m for all $t \geq 0$. Indeed, repeating the simulation using the same initial conditions for a large number of interaction ranges and plotting the corresponding value of minimum average predator distance (see (12)) results in Figure 9(b), which shows that $\rho^* = 10$ m for this particular initial condition and set of parameters.

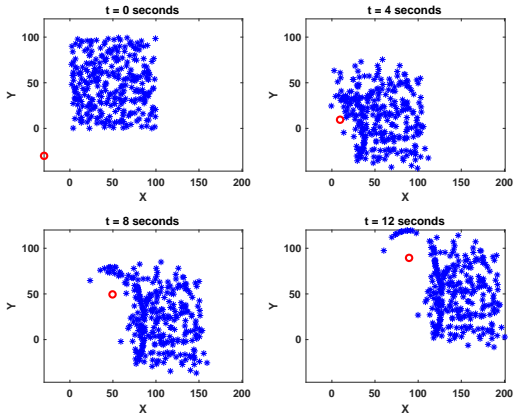

Figure 6: Predator represented by red "o". 3D Case. $\rho = 100\ m$.

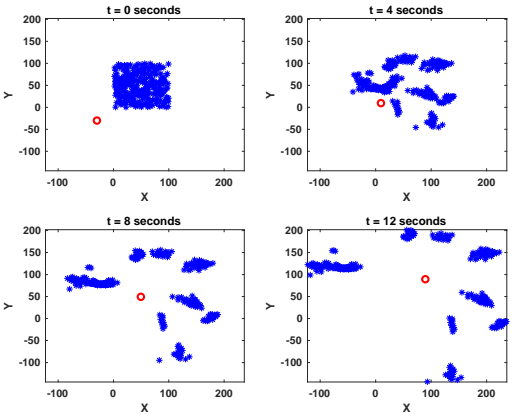

Figure 7: Predator represented by red "o". 3D Case. $\rho = 10\ m$.

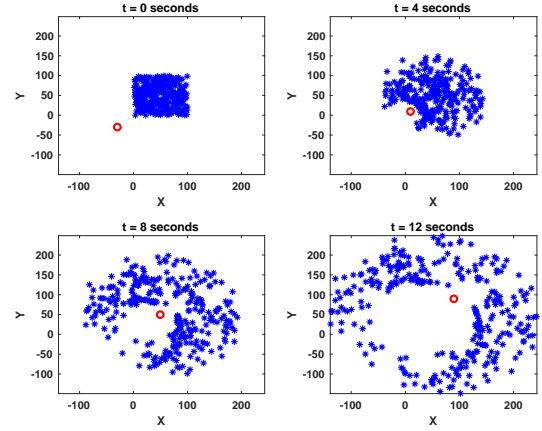

Figure 8: Predator represented by red "o". 3D Case. $\rho = 0\ m$.

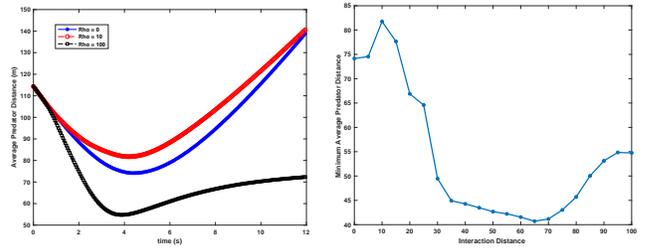

Figure 9: (a) Average Predator Distance for $\rho = 0, 10, 100\ m$. (b) Minimum Average Predator Distance as a function of $\rho$.

As mentioned earlier, instead of focusing on predator *average* distance as a measure of escape efficiency, one can consider predator *minimum* distance $\check{d}_\rho(t)$ from the group as a function of time as the yardstick for optimization – see (13). To this end, we show in Figure 10(a) the minimum predator distance function for the $\rho = 100, 10,$ and $0$ m cases, and display the minimizer of that function for a range of $\rho$s in Figure 10(b), demonstrating once again the power of small non-zero interaction ranges (in the span of 10 to 30 m ranges in this particular case).

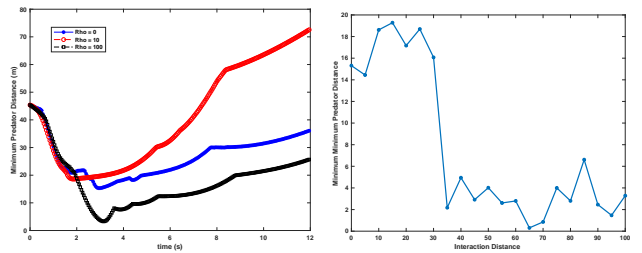

Figure 10: (a). Minimum Predator Distance for $\rho = 0, 10, 100\ m$. (b) Minimum Minimum Predator Distance as a function of $\rho$.

## VI. CONCLUSION

In this paper, we consider the problem of finding the optimal interaction range for consensus-driven multi-agent dynamic systems under attack by a single predator that maintains a constant velocity. It is shown that for both the 1D and 3D cases, there are a number of scenarios where the optimal range corresponds to forcing the underlying network to break up into a handful of disconnected graphs, each containing a subset of the agents. It is important to note that the results are largely based on numerical simulations, and are in fact dependent on a number of system parameters. For example, the reported results correspond to a predator approaching a flock somewhere near its center of mass. If instead the predator approached an "edge" of the flock, then breaking up into several clusters may no longer be the optimal solution for an efficient escape – full network connectivity may be more advantageous then. This point, together with an investigation of analytical solutions for the optimization problem posed, are the subject on ongoing research.

Additional future research extensions for the work reported in this paper can proceed in a number of directions. For example, one may consider a more sophisticated predator than the one assumed here by allowing the predator to dynamically update its velocity to increase its own efficiency and the probability of catching an agent as in classic pursuit problems. Furthermore, one may consider a multi-predator scenario, with the possibility of cooperation among the predators to maximize the probability of a hit.

Another potential future research direction is to focus on group dynamic behavior beyond simple consensus. For example, one could consider the classic Reynolds flocking algorithm [23] to represent agents' dynamics. Alternatively, one could study the escape efficiency problem when the underlying consensus algorithm is defined by a topological distance rather than a metric distance [24].

VII. ACKNOWLEDGEMENT

The author wishes to thank Professor Santiago Segarra and Dr. Lisa O'Bryan of Rice University for valuable discussions and for pointing out relevant papers in the opinion dynamics and biology literature.